\documentclass[11pt,a4paper]{iopart}
\usepackage{varioref}
\usepackage[dvips]{graphicx}
\usepackage[latin1]{inputenc}
\usepackage{iopams}

\begin{document}

\title{Strangeness production at SPS energies from NA49}

\author{Michael K. Mitrovski for the NA49 Collaboration\footnote[1] {Presented at Strangeness in Quark Matter 2006, Los Angeles, California, USA}}

\address{Institut f\"ur Kernphysik, J.W. Goethe-Universit\"at, 60438 Frankfurt, Germany}
\ead{Michael.Mitrovski@cern.ch}

\begin{abstract}
We present results on strange particle production from experiment NA49 in central and minimum bias Pb+Pb collisions in the beam energy range 20{\textit A} - 158{\textit A} GeV. New results on $\Xi$ production in central Pb+Pb collisions and on $\Lambda$, $\Xi$ production in minimum bias collisions are shown. Transverse mass spectra and rapidity distributions of strange particles at different energies are compared. The energy dependence of the particle yields and ratios is discussed. NA49 measurements of the $\Lambda$ and $\Xi$ enhancement factors are shown for the first time.  

\end{abstract}

\submitto{\JPG}

\section{Introduction}
The NA49 experiment has collected data on Pb+Pb collisions at beam energies between 20{\textit A} - 158{\textit A} GeV with the objective to search for the phase transition to a deconfined phase which is expected to occur in the early stage of the reactions if a sufficient energy density is achieved. \\
Several experimental signatures for a QGP have been proposed, among which strange particle yield measurements, in particular, are considered to be a powerful tool. An enhancement increasing with the strangeness content of the particle was predicted as a consequence of a QGP formation \cite{Rafelski_1} $-$ \cite{Rafelski_3}. \\
A non-monotonic energy dependence of the $K^{+}$/$\pi^{+}$ ratio with a sharp maximum close to 30{\textit A} GeV is observed in central Pb+Pb collisions. Within the statistical model of the early stage \cite{Marek_Paper_Early_stage}, this is interpreted as a signal of the onset of deconfinement, which causes a sharp change in the energy dependence of the strangeness to entropy ratio. This observation naturally motivates further study of the production of strange hadrons, in particular hyperons as a function of the beam energy. \\
In this report we show new NA49 results on strange particle production in central and minimum bias Pb+Pb collisions at all available SPS energies and compare them to previously shown data \cite{Marek_Proceedings} $-$ \cite{Proton_Paper} at other energies, as well as to model predictions.

\section{The NA49 experiment}

The NA49 detector \cite{NA49NIM} is a large acceptance hadron spectrometer at the CERN SPS, featuring four large volume TPCs as tracking detectors. Two of them are inside a magnetic field. The ionisation energy loss (dE/dx) measurements in the TPCs are used for particle identification. Central collisions were selected on the basis of the energy of the projectile spectator nucleons measured by a downstream calorimeter.

\section{Energy Dependence of Transverse Mass Spectra}

\begin{figure}[h!]
\begin{center}
\includegraphics[scale=0.4]{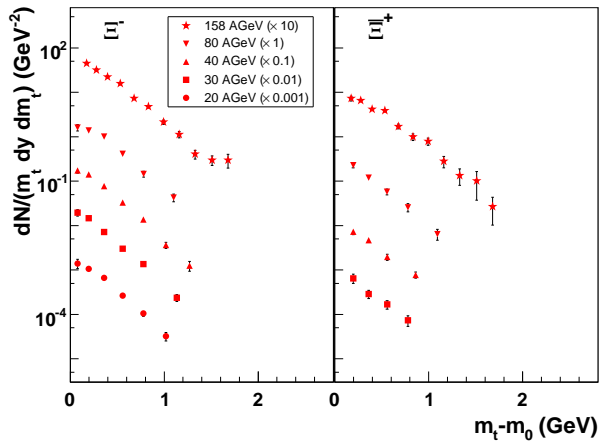} 
\end{center} 
\caption{Transverse mass spectra of $\Xi^{-}$ (left) and $\bar{\Xi}^{+}$ (right) in central Pb+Pb (7~\% most central at 20{\textit A} $-$ 80{\textit A} GeV, 10 \% most central at 158{\textit A} GeV) collisions at midrapidity. The errors shown are statistical only.}  
\label{fig:mtXi}
\end{figure}

\begin{figure}[h!]
\begin{center}
\includegraphics[scale=0.37]{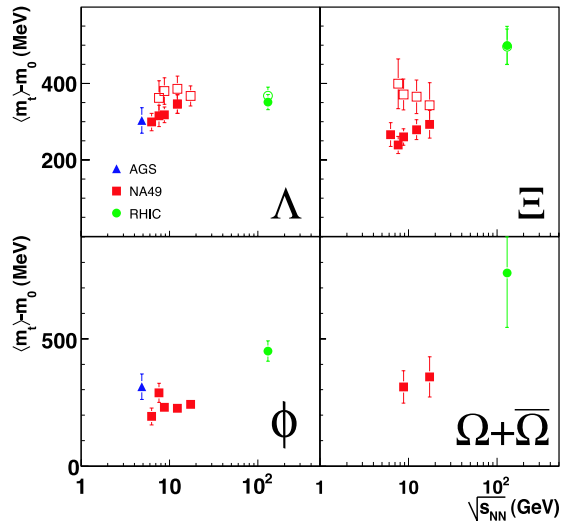} 
\end{center} 
\caption{Energy dependence of the mean transverse mass $\langle m_{t} \rangle$ $-$ $m_{0}$ of $\Lambda$, $\phi$, $\Xi$ and $\Omega$ + $\bar{\Omega}$ in central Pb+Pb/Au+Au collisions from AGS \cite{E896Lambda}, \cite{E917_Phi} to RHIC \cite{STAR1} energies. Open symbols indicate the anti-particles.}  
\label{fig:Meanmt}
\end{figure}

Fig.~\ref{fig:mtXi} shows new results on transverse mass spectra of $\Xi^{-}$ (left) and $\bar{\Xi}^{+}$ (right) integrated over the rapidity range -0.5 $<$ y $<$ +0.5 in central Pb+Pb collisions at different beam energies. The shape of the spectra is approximately exponential in $m_{t}$-$m_{0}$ with an inverse slope parameter of about $T(\Xi^{-})$ $\approx$  240 MeV and $T(\bar{\Xi}^{+})$ $\approx$ 280 MeV, and exhibits no significant energy dependence. \\
 \\
In Fig.~\ref{fig:Meanmt} the energy dependence of the mean transverse mass, $\langle m_{t} \rangle$ $-$ $m_{0}$, is shown for $\Lambda$, $\phi$, $\Xi$ and $\Omega$ + $\bar{\Omega}$. The mean transverse mass was calculated from the measured spectra and using a fitted function extrapolated to full $m_{t}$ range. Different extrapolation methods have been tested to estimate systematic error. The resulting extrapolation factors are small. \\
 \\
If energy density and hence pressure increase with beam energy, an increase of transverse expansion is expected. Assuming that the strength of the transverse expansion is reflected in the mean transverse mass, the variable $\langle m_{t} \rangle$ $-$ $m_{0}$ is expected to rise with $\sqrt{s_{NN}}$. Previously we have reviewed the energy dependence of $\langle m_{t} \rangle$ $-$ $m_{0}$ for pions, kaons and protons \cite{BLUME} - \cite{E866_E917_Part}. At AGS energies these particles show a strong increase of the mean transverse mass with the beam energy. However, the mean transverse masses show a weak energy dependence at SPS energies. The weak dependence has been attributed \cite{MAREK_GOREN} to the constant pressure and temperature when a mixed phase is reached during the dynamical evolution of the colliding system in the early stage of the reaction. Fig.~\ref{fig:Meanmt} shows new results on $\langle m_{t} \rangle$ $-$ $m_{0}$ for heavy strange hadrons. Also for these particles no significant energy dependence is observed at SPS energies. Measurements of multiply strange hadrons at AGS energies are needed to fully study their energy dependence. 

\section{Energy Dependence of Particle Yields}

\begin{figure}[h!]
\begin{center}
\includegraphics[scale=0.38]{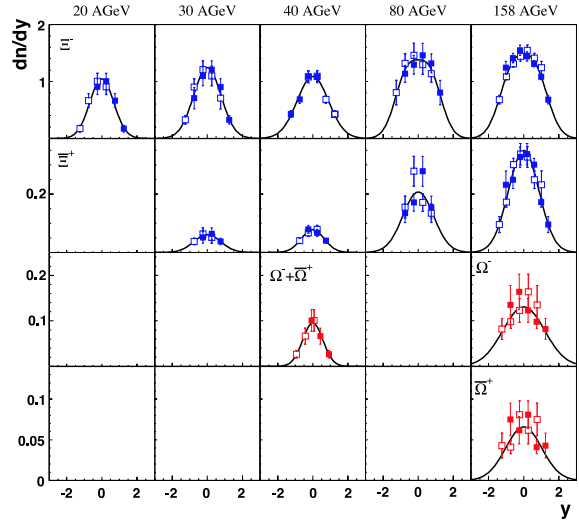} 
\end{center} 
\caption{The rapidity spectra of $\Xi$ and $\Omega$ produced in central Pb+Pb collisions (7 \% most central at 20{\textit A} $-$ 80{\textit A} GeV, 10 \% ($\Xi$) and 23.5 \% ($\Omega$) most central at 158{\textit A} GeV). The closed symbols indicate measured points, open points are reflected with respect to midrapidity. The solid lines represent fits with a single Gaussian. The errors shown are statistical only.}
\label{fig:rap}
\end{figure}

\begin{figure}[h!]
\begin{center}
\includegraphics[scale=0.33]{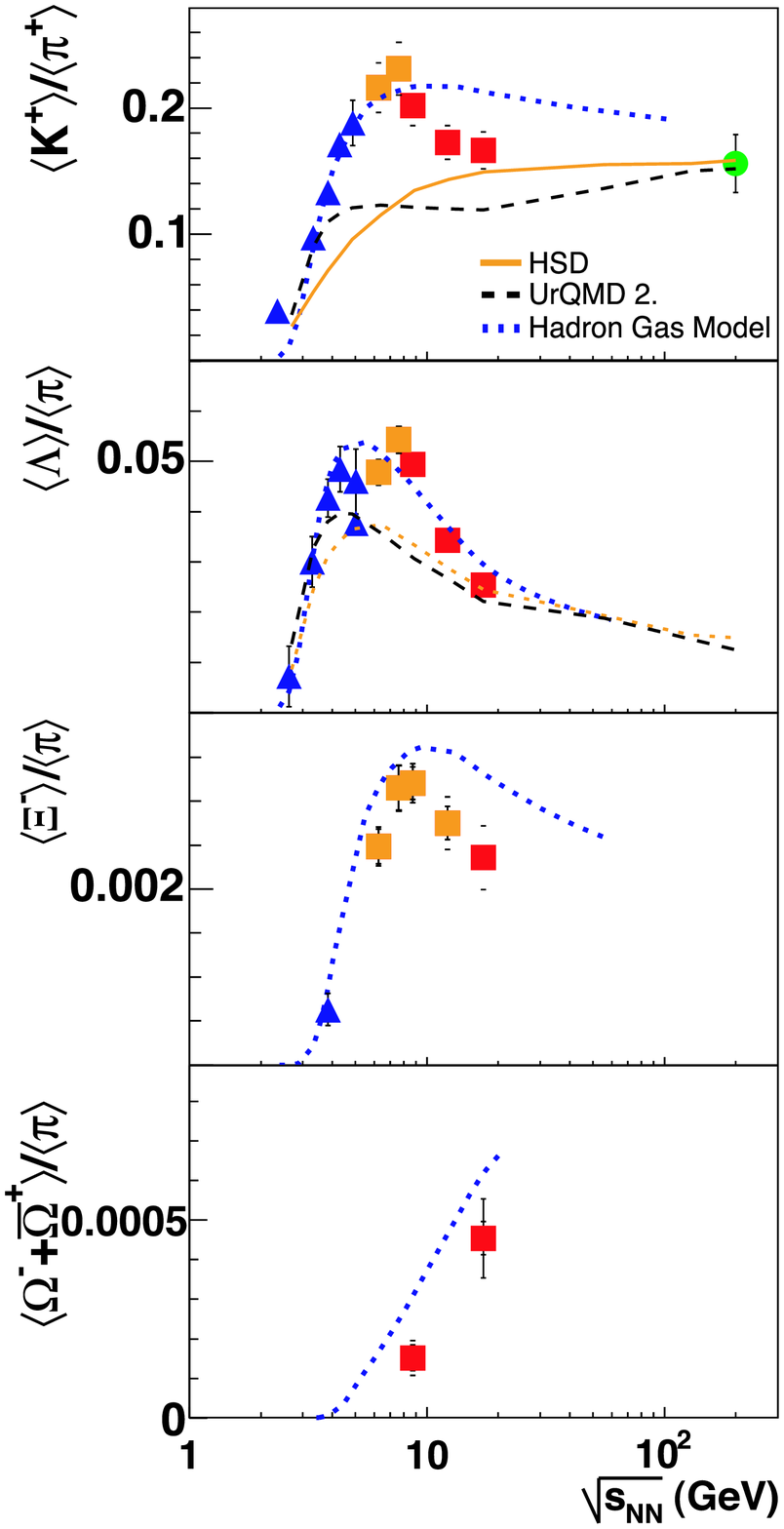} 
\includegraphics[scale=0.33]{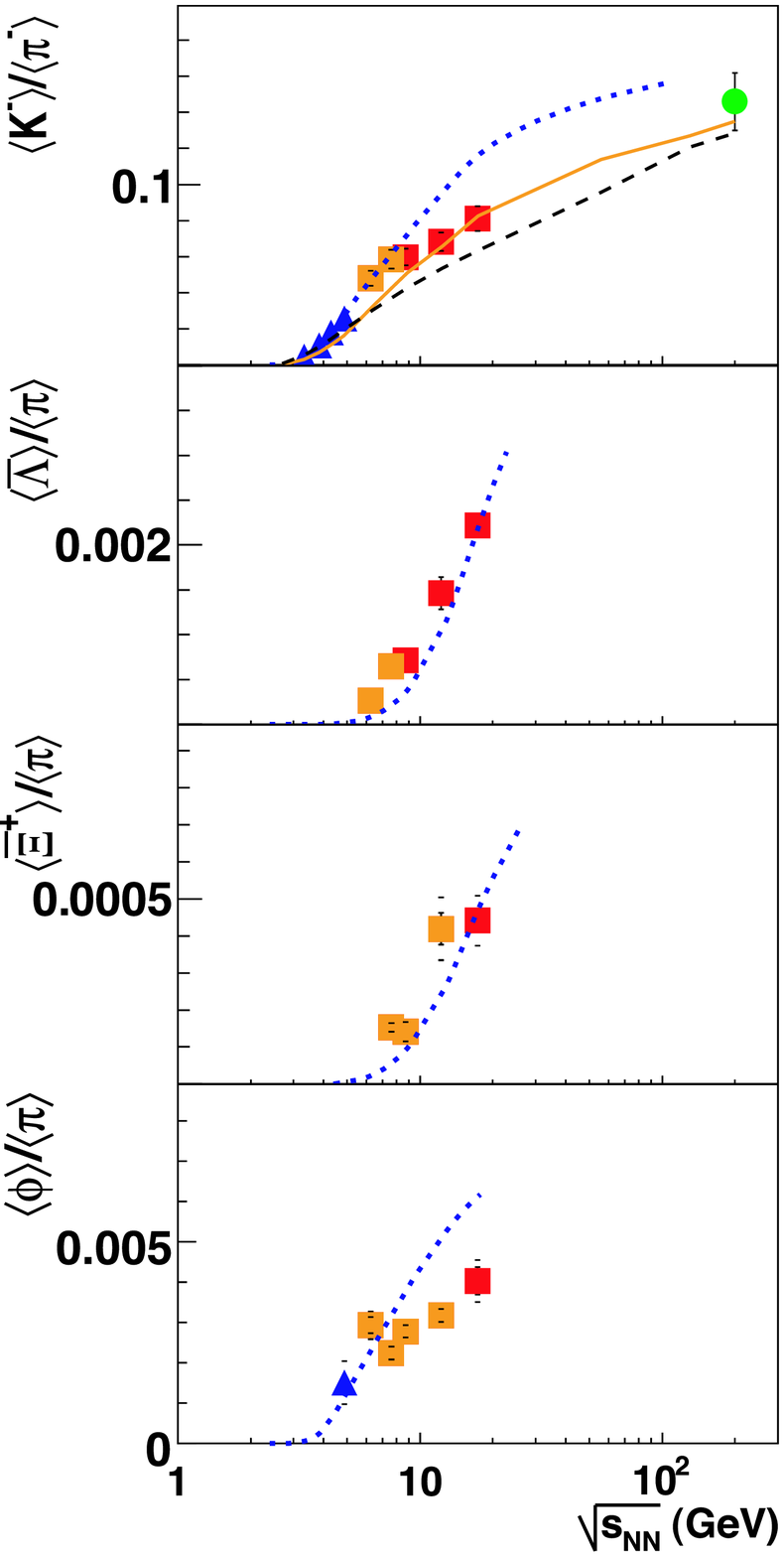} 
\end{center} 
\caption{The energy dependence of the total yields of strange hadrons, normalized to the total pion yields ($\langle$$\pi$$\rangle$ = 1.5 ($\langle$$\pi^{-}$$\rangle$ + $\langle$$\pi^{+}$$\rangle$), in central Pb+Pb/Au+Au collisions. The data are compared to string hadronic models \cite{BRATKOV} (UrQMD 2.0: dashed, HSD: solid) and statistical hadron gas model \cite{CLEYMANS} (assuming strangeness in full equilibrium: dotted). }  
\label{fig:particle/pi_yield_4pi_ratio}
\end{figure}

\begin{figure}[h!]
\begin{center}
\includegraphics[scale=0.4]{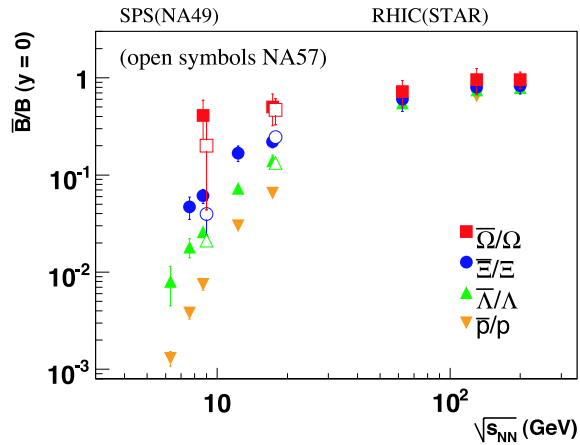} 
\end{center} 
\caption{The antibaryon/baryon ratio ($\bar{B}$/$B$) at midrapidity from SPS to RHIC energies. The errors shown are statistical only.}  
\label{fig:AntiBaryon/Baryon_ratio}
\end{figure}

\begin{figure}[h!]
\begin{center}
\includegraphics[scale=0.31]{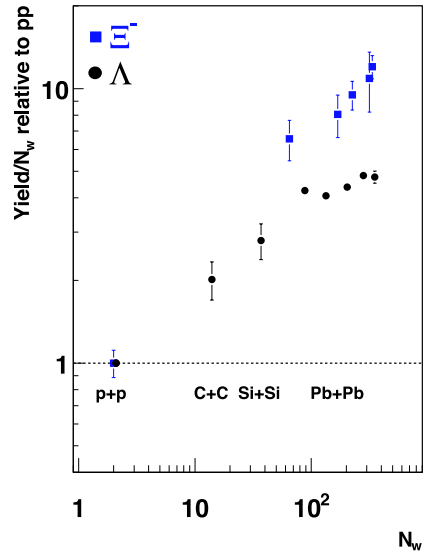} 
\end{center} 
\caption{Hyperon enhancement as a function of the number of wounded nucleons at 158{\textit A} GeV. The errors shown are statistical only.}  
\label{fig:strangeness_enhancement}
\end{figure}

The NA49 experiment features a large acceptance in the forward hemisphere allowing for measurements of rapidity spectra from midrapidity up to almost beam rapidity. The rapidity spectra for $\Xi$ and $\Omega$ in central Pb+Pb collisions are shown in Fig.~\ref{fig:rap}. They are parametrized by a single Gaussian function. NA49 results on spectra of $\pi^{-}$, $K^{+}$, $K^{-}$, $\phi$, $\Lambda$, and $\bar{\Lambda}$ were presented earlier \cite{BLUME}. \\
\begin{figure}[h!]
\begin{center}
\includegraphics[scale=0.55]{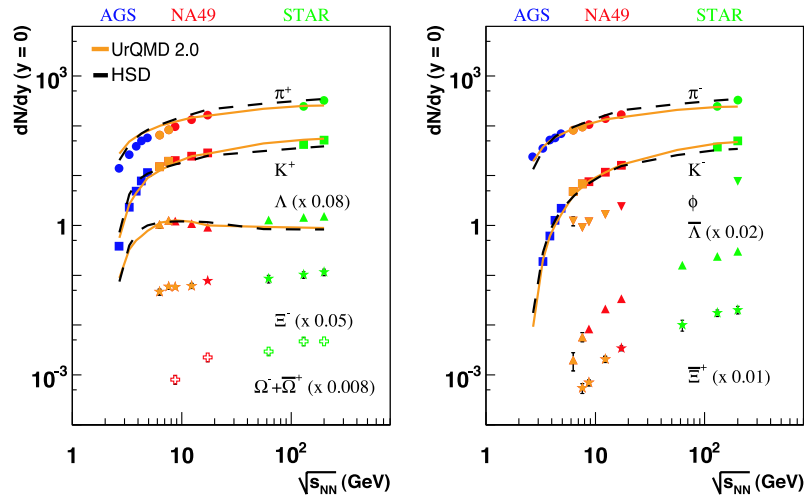} 
\end{center} 
\caption{Midrapidity yields from AGS to RHIC energies. The data are compared to string-hadronic models \cite{BRATKOV} (UrQMD 2.0: dashed, HSD: solid). The errors shown are statistical only.}  
\label{fig:MidYield}
\end{figure} \\
NA49 has measured different particle species at midrapidity in 20{\textit A}, 30{\textit A}, 40{\textit A}, 80{\textit A}, and 158{\textit A} GeV central Pb+Pb collisions. Results on yields at midrapidity in central Au+Au collisions at $\sqrt{s_{NN}}$ = 62.4, 130 and 200 GeV have been published by the STAR collaboration at RHIC \cite{SPELTZ}. A compilation of these data is shown in Fig.~\ref{fig:MidYield}, together with AGS data. \\
\begin{figure}[h!]
\begin{center}
\includegraphics[scale=0.33]{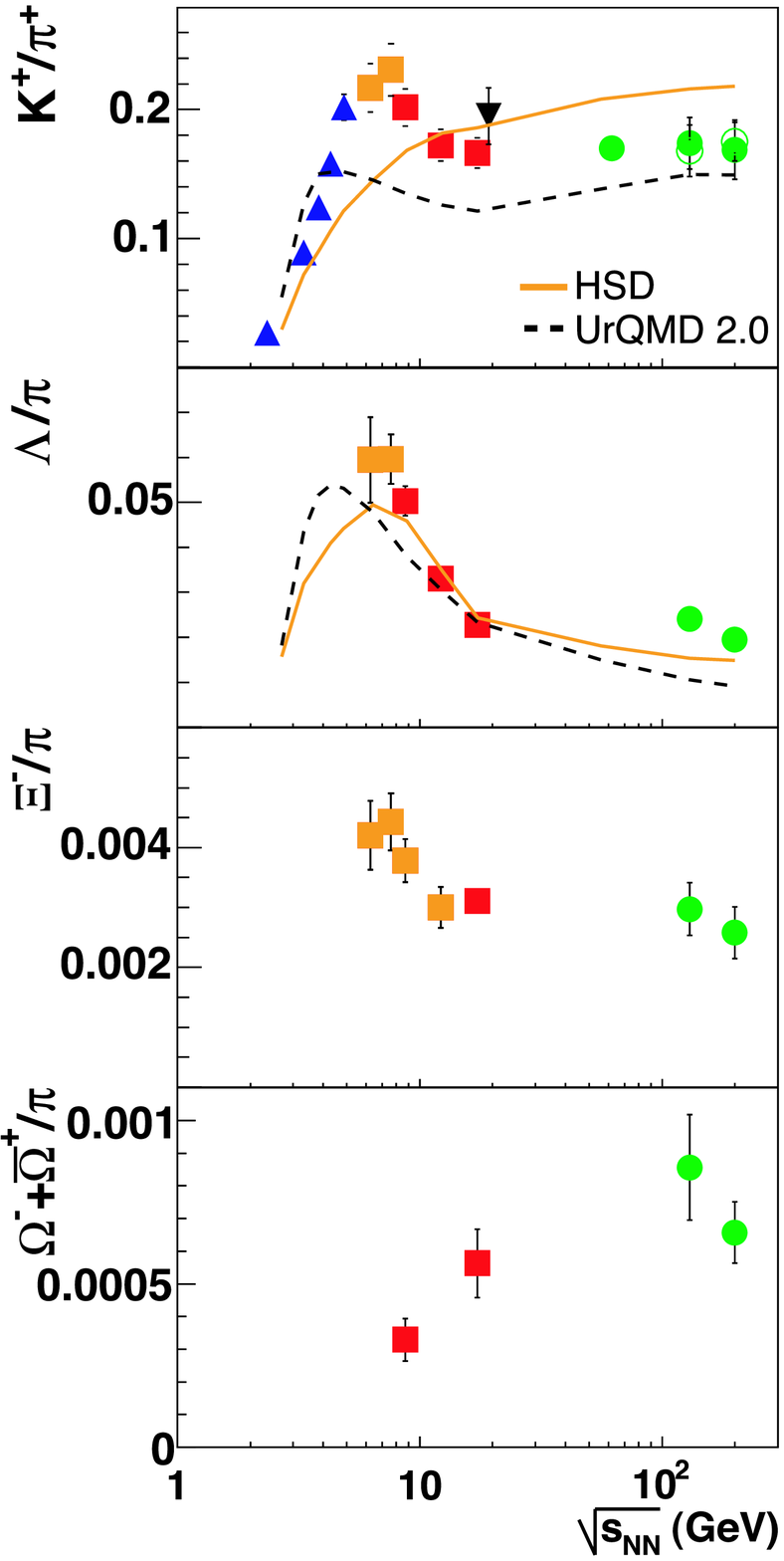} 
\includegraphics[scale=0.33]{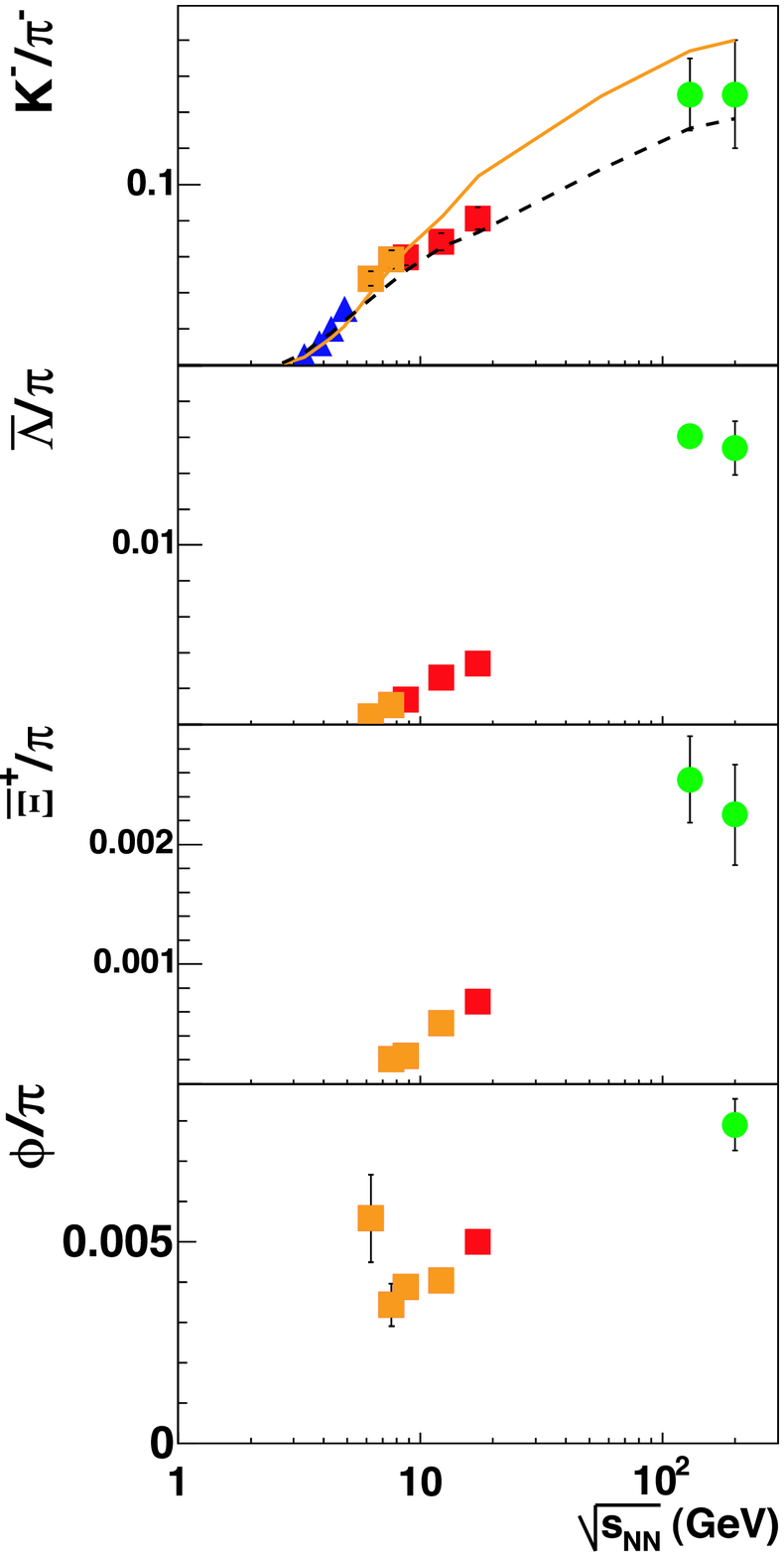} 
\end{center} 
\caption{The energy dependence of the midrapidity yields of strange hadrons, normalized to the midrapidity pion yields ($\pi$ = 1.5 ($\pi^{-}$ + $\pi^{+}$)), in central Pb+Pb/Au+Au collisions. The data are compared to string hadronic models \cite{BRATKOV} (UrQMD 2.0: dashed, HSD: solid).}  
\label{fig:particle/pi_yield_midrap_ratio}
\end{figure} \\
One observes a relatively weak energy dependence of the yield of hyperons starting from SPS to RHIC energies, while a strong increase is seen for the anti-hyperons. The data on $\Lambda$ and kaon yields are compared to string-hadronic models \cite{BRATKOV}. Both models follow the trend of the measured data, but they fail to reproduce the data quantitatively. Predictions for multi-strange hyperons and the $\phi$-meson are still not available.  \\ \\
Figs.~\ref{fig:particle/pi_yield_midrap_ratio} and \ref{fig:particle/pi_yield_4pi_ratio} show the energy dependence of relative strange particle yields. The ratio of the midrapidity and total yields normalized to the respective pion yields are compared to model calculations. While $\langle$$K^{-}$$\rangle$/$\langle$$\pi^{-}$$\rangle$, $\langle$$\bar{\Lambda}$$\rangle$/$\langle$$\pi$$\rangle$, $\langle$$\bar{\Xi}^{+}$$\rangle$/$\langle$$\pi$$\rangle$, $\langle$$\Omega^{-}+\bar{\Omega}^{+}$$\rangle$/$\langle$$\pi$$\rangle$ and $\langle$$\phi$$\rangle$/$\langle$$\pi$$\rangle$ ratios rise continuously with energy, a distinct maximum is visible in the energy dependence of the ratios $\langle$$K^{+}$$\rangle$/$\langle$$\pi^{+}$$\rangle$, $\langle$$\Lambda$$\rangle$/$\langle$$\pi$$\rangle$ and $\langle$$\Xi^{-}$$\rangle$/$\langle$$\pi$$\rangle$. The same structure is seen for the midrapidity ratios. Hadron gas \cite{CLEYMANS} (only available for 4$\pi$ multiplicity ratios) and microscopic models \cite{BRATKOV} do not provide a proper description of the structures observed in the data. On the other hand, this feature can be understood in a scenario where the onset of deconfinement is reached around 30{\textit A} GeV as proposed in the statistical model of the early stage \cite{Marek_Paper_Early_stage}. \\
 \\
The NA49 antibaryon/baryon ratios are shown in Fig.~\ref{fig:AntiBaryon/Baryon_ratio} as a function of the beam energy. The ratios increase with increasing strangeness content of the hyperons at a given energy. The energy dependence of the ratio for multiply strange hyperons is much weaker than that for Protons and $\Lambda$. Because the net baryon density at midrapidity is a strong function of the beam energy, the antibaryon to baryon ratio strongly depends on the number of valence up and down quarks in the hadron. For $\Omega$ with no valence up/down quarks the energy dependence of the ratio is the weakest. The open symbols represent measurements from the NA57 Collaboration \cite{ANTINORI}. While the particle ratios are in good agreement with the NA49 data, the absolute yields of NA57 tend to be systematically higher by about 30 \%, in central Pb+Pb collisions both at 40{\textit A} and 158{\textit A} GeV \cite{ELIA}. 
 \\
Fig.~\ref{fig:strangeness_enhancement} shows results on system size and centrality dependence of hyperon yields at midrapidity at 158{\textit A} GeV. Having measured particle yields for p+p, C+C, Si+Si and Pb+Pb collisions, one can determine a hyperon enhancement as : \\
\begin{displaymath}
E = \left( \frac{Yield}{N_{w}} \right)_{A+A}  / \left( \frac{Yield}{N_{w}} \right)_{p+p},
\end{displaymath} 
where $N_{w}$ is the number of wounded nucleons calculated within the Glauber model \cite{GLAUBER}. A significant enhancement for the $\Xi^{-}$ is observed at 158{\textit A} GeV when going from p+p to Pb+Pb. For the $\Lambda$ hyperon the enhancement is not as strong as for the $\Xi^{-}$ hyperon and the centrality dependence is slightly weaker. Together with the results for C+C and Si+Si \cite{System_Size} a complete coverage of the system size range from $N_{w}$ $>$ 14 is possible.  \\
 \\
The NA57 collaboration \cite{ANTINORI} observes the same trend (not shown) but instead of p+p a p+Be reference is used. Since already in p+A reactions a slight enhancement of strange particle production is observed \cite{Xi_Tanja}, the enhancement relative to p+Be is less. \\ \\
Particle yields up to $p_{t}$ = 4 GeV/c have been measured in central Pb+Pb collisions. The baryon/meson ratios qualitatively show the same $p_{t}$ dependence as at RHIC. As well, the nuclear modification factor $R_{CP}$ is extracted and compared to pQCD calculations \cite{SCHUSTER}. \\ \\
The elliptic flow of $\Lambda$ hyperons has been measured in semi-central Pb+Pb collisions at 158{\textit A} GeV. The standard method of correlating particles with an event plane has been used. The elliptic flow of $\Lambda$ particles increases linearly up to 2 GeV/c. This increase is weaker at SPS than at RHIC energies, partly due to different centrality selection. A linear increase of $v_{2}$ is also observed for other particles ($\pi$,proton) as a function of $p_{t}$ in mid-central collisions at 158{\textit A} GeV \cite{KIKOLA}. 

\section*{Acknowledgments} 
Acknowledgements: This work was supported by the US Department of Energy
Grant DE-FG03-97ER41020/A000,
the Bundesministerium fur Bildung und Forschung (06F137), Germany, 
the Virtual Institute VI-146 of Helmholtz Gemeinschaft, Germany,
Fellowship of the German Academic Exchange Service (DAAD) for Ph.D.-research studies, Germany,
the Polish State Committee for Scientific Research (1 P03B 097 29, 1 PO3B 121 29,  2 P03B 04123), 
the Hungarian Scientific Research Foundation (T032648, T032293, T043514),
the Hungarian National Science Foundation, OTKA, (F034707),
the Polish-German Foundation, the Korea Research Foundation Grant (KRF-2003-070-C00015) and the Bulgarian National Science Fund (Ph-09/05).

\end{document}